\documentclass{comnet}

\usepackage{graphicx}
\usepackage{amsmath}
\usepackage{mathrsfs}
\usepackage{color}

\begin{document}

\title{A Generative Hypergraph Model for Double Heterogeneity}

\shorttitle{A Generative Hypergraph Model for Double Heterogeneity} 
\shortauthorlist{Zhao Li, Jing Zhang, Guozhong Zheng et al.} 

\author{
    \name{Zhao Li , Jing Zhang  , Guozhong Zheng and  Li Chen$^\dagger$}
    \address{School of Physics and Information Technology, Shaanxi Normal University, \\Xi’an 710062, P. R. China
    \email{$^\dagger$Corresponding author. Email: chenl@snnu.edu.cn}}
    \and
    \name{Jiqiang Zhang}
    \address{School of Physics and Electronic-Electrical Engineering, Ningxia University, Yinchuan 750021, P. R. China}
    \name{Weiran Cai}
    \address{School of Computer Science, Soochow University, Suzhou 215006, P. R. China}
}


\maketitle

\begin{abstract}
{While network science has become an indispensable tool for studying complex systems, 
the conventional use of pairwise links often shows limitations in describing high-order interactions properly. 
Hypergraphs, where each edge can connect more than two nodes, have thus become a new paradigm in network science. Yet, we are still in lack of models linking network growth and hyperedge expansion, both of which are commonly observable in the real world. Here, we propose a generative hypergraph model by employing the preferential attachment mechanism in both nodes and hyperedge formation. The model can produce bi-heterogeneity, exhibiting scale-free distributions in both hyperdegree and hyperedge size. We provide a mean-field treatment that gives the expression of the two scaling exponents, which agree with the numerical simulations. Our model may help to understand the networked systems showing both types of heterogeneity and facilitate the study of complex dynamics thereon.}
{Hypergraph; Scale-free distribution; Power law; Complex networks}
\\
\end{abstract}
\textbf{Glossary of Terms}

\emph{Hypergraph}: a hypergraph is an extension of a graph, where an edge can connect any number of vertices/nodes. Formally, a hypergraph is defined as $H=(V,E)$, where $V=\{v_1, v_2,...,v_n\}$ is a finite set of nodes, and $E=\{e_1, e_2,..., e_m\}$ is a group of subsets of elements in $V$, corresponding to the hyperedge set. When all hyperedge include only two nodes (i.e. $|e_i| = 2$), the hypergraph is recovered into a graph. In this work, we use this term interchangeable with hypernetwork.

\emph{u-uniform Hypergraph}: a hypergraph where all its hyperedges are of the same size,  $|e_i| = u, u\in\{2, 3, ...\}$.


\emph{Hyperdegree}: the number of hyperedges that include node $i$, denoted as $d_H(i)$.

\emph{Hyperedge degree}: the number of hyperedges that connect hyperedge $e_i$, denoted as $d_{Hd}(i)$. Two hyperedges $e_{i,j}$ are connected if they include at least one same node, i.e. $e_i \cap e_j \neq \phi$. 

\emph{Size of hyperedge}: the number of nodes that a hyperedge $e_i$ includes, denoted as $S(i)$.

\section{Introduction}
A network is a simplest abstraction of a complex system~\cite{newman2010}, where the components of the system are represented as nodes, which are usually linked by edges depicting interactions or relationships between two involved components. Since the seminal work of Watts-Strogatz model~\cite{Watts1998} and the Barab\'asi-Albert model~\cite{ba1999Emergence}, network science has rapidly developed into a powerful paradigm for studying complex natural and socio-economic systems ~\cite{Leo2021,Strogatz2001,Albert2002,Newman2003,Albert2004}. 

However, some scenarios often arise where interactions go beyond pairwise as in the traditional network paradigm, such as in scientific collaborations~\cite{patania2017}, neuronal activities~\cite{Giusti2015}, biological processed~\cite{steffen2009}, social networks~\cite{Giulia2021} and cell-to-cell communications~\cite{Ritz2014}. There, more than two components can be simultaneously engaged in a group interaction, also termed as higher-order interactions. In this regard, hypergraphs or hypernetworks were coined where the components, still represented by a set of nodes, can be linked by hyperedges that are allowed to connect multiple nodes~\cite{Berge1973}. This extension has proved to be more powerful in describing many complex systems in reality, where the traditional pairwise network formulation can be taken as special cases of hypergraphs. Owing to its efficacy, the concept of hypergraphs has boomed to a widely adopted framework, which is undoubtedly helpful for depicting numerous real circumstances~\cite{Lambiotte2019networks, Battiston2020networks, Boccaletti2023structure}. 

To understand the features of hypergraphs, a useful way is to build generative models that reproduce the properties of real networks. Prominent examples include Erd\H{o}s-R\'enyi (ER) random network model, Watts-Strogatz model~\cite{Watts1998}, and Barab\'asi-Albert (BA) model~\cite{ba1999Emergence} that have been leveraged to understand the basic pictures within the classic network framework. In this regard, there are a few works that have started to model the realistic instances within the hypergraph framework.
One example is the folksonomy, where the tripartite structure of users, resources, and tags are often seen, and a random hypergraph model is proposed and results are compared with data from Flickr~\cite{ghoshal2009}, this website together with the bookmarking site CiteULike share similar properties with many previously studied complex networks~\cite{Zlatic2009}.
Within the context of knowledge generation and diffusion, the hypernetwork is also found more proper for description, and two models are proposed by integrating the hypernetwork structure and the knowledge generation processes~\cite{Liu2014}.
From the hypergraph perspective, Vazquez constructs a statistical model to solve the population stratification problem, which is tested upon phenotypic or genetic information~\cite{alixei2008}. 

In the meantime, there are a few work devoting to building minimal generative model for hypergraphs, and focusing on the statistical properties. 
For example, a class of evolving models for hypergraphs are proposed ~\cite{wang2010, Hu2013, WU2014, guo2014a}, where by adopting the growth and linear preferential attachment the degree or hyperdegree distribution renders scale-free property.  
In addition, nonlinear preferential attachment is also adopted~\cite{guo2014b}, where the number of newly added nodes is also draw from a distribution, thus resulting a nonuniform hypergraph. A combination of linear preferential attachment and/or nonpreferential mechanism is also developed in~\cite{Kovalenko2021growing}.
Ref.~\cite{lu2021} propose a hypergraph model inspired by ER random graphs, where the node hyperdegree distribution obeys a Poisson distribution; they also build bilayer hypergraph model with ER or BA property in each layer.
Besides, there are also some efforts for generative hypergraph construction based upon different preferential mechanisms aiming to model the real-world data~\cite{Yang2014,wang2020,hufeng2021,sun2020,zhou2020,guojinli2014}. 

However, a common feature of most proposed hypergraph models is the homogeneous distribution in the hyperedge size. This is in sharp contrast with the observations in reality, where the sizes of hypergraphs are often heterogeneous. For example, in social hypernetworks where nodes represent the individuals while hyperedges link groups such as classes, companies, and even towns, the hyperedges generally vary in their sizes. The sizes of hyperedges (i.e. the firm size) in company hypernetworks are found to follow a power-law distribution~\cite{Axtell2001zipf}. Similar observations were also be made for the urban sizes~\cite{Newman2005} , where each town or city can be taken as a hyperedge. 
We also give two examples showing the heterogenous property in the SM~\cite{SM}. One is the number of publications distribution for researchers in four scientific journals, demonstrating power law distributions in the hyperdegree. The other is urban population distribution in China in 2000, showing the heterogeneity in the hyperedge.
It is hence natural to develop a hypergraph model that is able to produce skewed distributions in both hyperdegree and hyperedge size, which is able to embrace the empirically observed two heterogeneities in a unified framework. 

In this work, we develop a double heterogeneity hypergraph model. 
By incorporating a continual growth mechanism with preferential attachment, in both nodes or hyperedges, such a simple model can generate scale-free graphs exhibiting the signature skewed distributions in both hyperdegrees and hyperedges. We develop a mean-field theory that provides the exact expressions of the two scaling exponents, which are in good agreement with the numerical simulations. 

The reminder of this paper is organized as follows. In Sec.~\ref{model}, we introduce our model. The numerical results are given in Sec.~\ref{simulation}, with a theoretical treatment to derive the two associated power exponents in Sec.~\ref{theory}. Sec.~\ref{comparison} provides the comparison between the numerical results and the theoretic prediction. Conclusions and discussions will be given in Sec.~\ref{conclusion}. 

\begin{figure}[tbp]
	\centering
	\includegraphics[width = 0.7\textwidth]{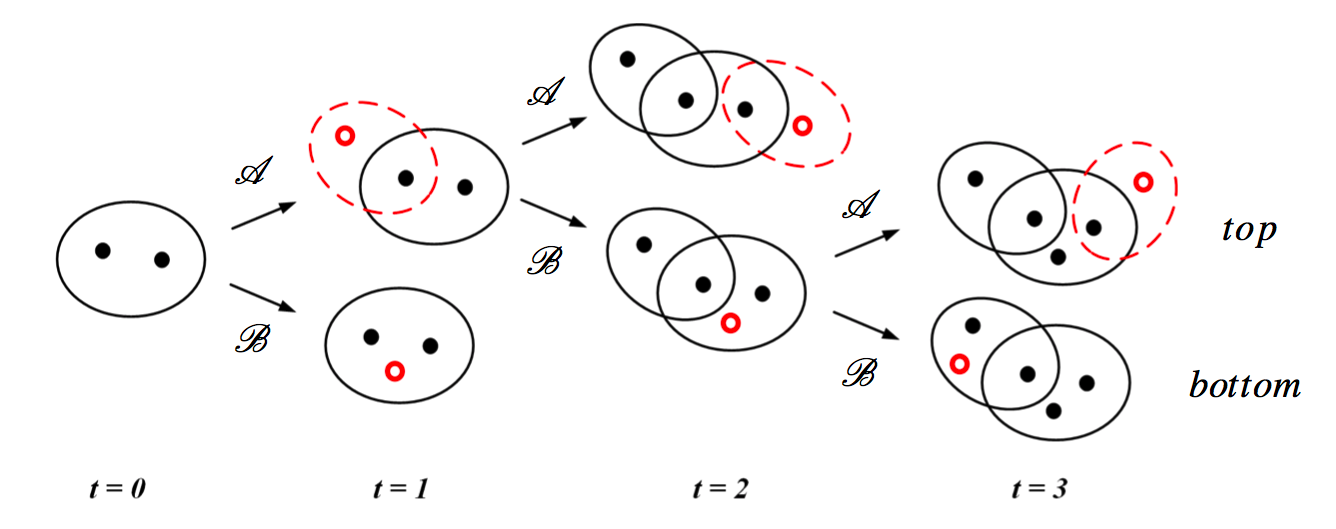}
	\caption{A schematic description of our DHH model with $m_0 = 2$, $m_{new} = 1$ and $m_{old} = 1$. The arrow  represents the growth process through either the occurrence of hyperedge formation labeled as $\mathscr{A}$ or the hyperedge expansion labeled as $\mathscr{B}$. When executing $\mathscr{A} - \mathscr{B} - \mathscr{A}$ in the first three steps respectively, the \emph{top} network can be obtained; when executing $\mathscr{A} - \mathscr{B} - \mathscr{B}$, the \emph{bottom} network is generated.} 
	\label{DHHscheme}
\end{figure} 

\section{Double heterogeneous hypergraph model}\label{model}
The generative model of the double heterogeneous hypergraph (DHH) incorporates two most commonly adopted ingredients as proposed in the BA model~\cite{ba1999Emergence}:  \emph{growth} and \emph{preferential attachment}. 
The former is due to the continuous expansion of networked systems, such as the growth of the internet, company networks, and the social networks etc; the latter is based on the observations that the newly added node is more likely to connect the already highly connected nodes or hyperedges. A simple example is the job application, where the applicants generally tend to join the big companies for higher salaries and a better platform. Here, the difference from the BA model is that the preferential attachment is engaged in both node degrees and hyperedges in our model rather than merely the degrees in BA model. 

The scheme of construction of DHH is illustrated in Fig.~\ref{DHHscheme}. The procedure is as follows: 
\begin{enumerate}
\item Initially, $m_0$ nodes form a hyperedge;
\item $m_{new}$ new node are added in batch, with a probability $p$, which form a new hyperedge by connecting to $m_{old}$ existing nodes. The probability of the existing node $i$ to be chosen follows $\Pi(d_H(i)) = d_H(i) / \sum_{j} d_H(j)$, where $d_H(i)$ is the hyperdegree of node $i$ and $\sum_{j} d_H(j)$ is the sum over all existing nodes; 
\item Otherwise, the newly added nodes join one of the existing hyperedges without forming a new hyperedge. The selection probability of an existing hyperedge is $\Pi (S(i)) = S(i)/\sum_{j} S(j)$, where $S(i)$ is the size of the hyperedge $i$  and $\sum_{j} S(j)$ represents the sum over all existing hyperedges.
\item Step 2-3 repeat until the hypergraph size reaches a preset size $N$.
\end{enumerate}

Fig.~\ref{DHHscheme} shows an example of network growth with $m_0=2$ and $m_{new}=m_{old}=1$. Note that, in this case, our model recovers to the classic BA network model when $p=1$, because there is no growth in the hyperedges, all edges including only two nodes, and the preferential attachment acts exclusively on the degrees. For arbitrary values of $m_{new}$ and $m_{old}$ in this extreme, the model recovers to a $u$-uniform hypergraph, where $u=m_{new}+m_{old}$. 
In the other extreme case with $p=0$, there is only one hyperedge that includes all nodes. Generally, we are interested in cases in between (i.e., $0<p<1$), where both the number of hyperedges and their sizes evolve.

\begin{figure}[htbp]
    \centering
    \includegraphics[width = 0.85\textwidth]{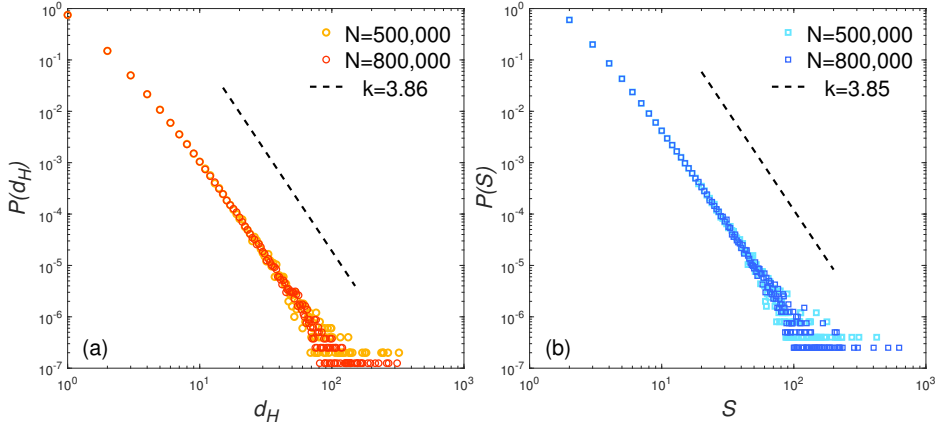}
    \caption{Distribution of hyperdegree and hyperedge size in our DHH model for two different sizes $N$, with $p=0.5$. 
    (a) The hyperdegree distributions for two system size ($N=5\times10^5 $ and $8\times10^5 $). The slope of the dashed line is $3.86$. 
    (b) The size distribution of hyperedges for the same two sizes. The slope of the dashed line is $3.85$. Parameters: $m_0=2$, $m_{new}=m_{old}=1$.
    }
    \label{powerlaw}
\end{figure} 

\begin{figure}[htbp]
    \centering
    \includegraphics[width = 0.85\textwidth]{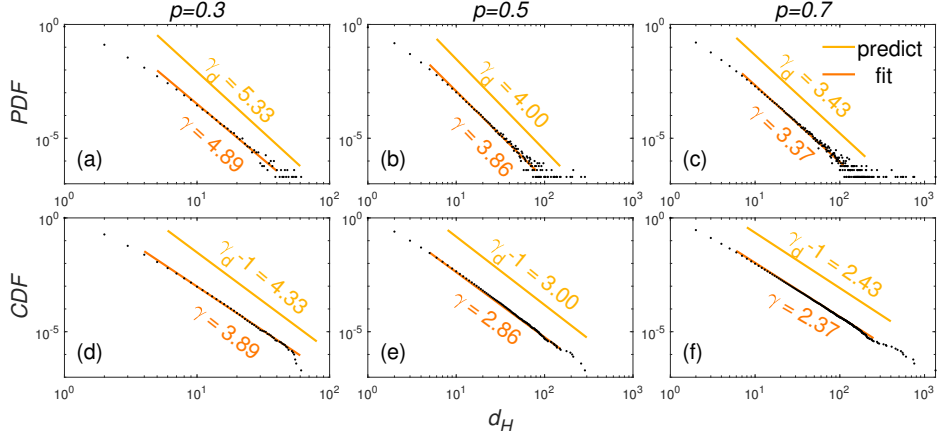}
    \caption{Hyperdegree distributions. (a-c) are the PDF of node hyperdegree for $ p = 0.3$, $0.5$, $0.7$, respectively. (d-f) are CDF of the same data as (a-c). The red line are the fitted power law by the least squares method, while the orange are the power law predicted by our mean-field theory given by Eq.~(\ref{gamma_d}). Parameters: $m_0=2$, $m_{new}=m_{old}=1$, and $N=5\times 10^5$. 
    }
    \label{hyperdegree}
\end{figure} 

\section{Simulation results}\label{simulation}
Following the description of the model in Sec.~\ref{model}, we first present simulation results with $m_0 = 2$, $m_{new}=m_{old}=1$ for simplicity. Bear in mind that the coevolving process includes the hyperedge formation with probability $p$ and the hyperedge expansion with $1-p$. After $t$ timesteps, the model yields a network of size $N_t=m_{new}t+m_0$, and $pt+1$ hyperedges on average. 

\begin{figure}[htbp]
    \centering
    \includegraphics[width = 0.85\textwidth]{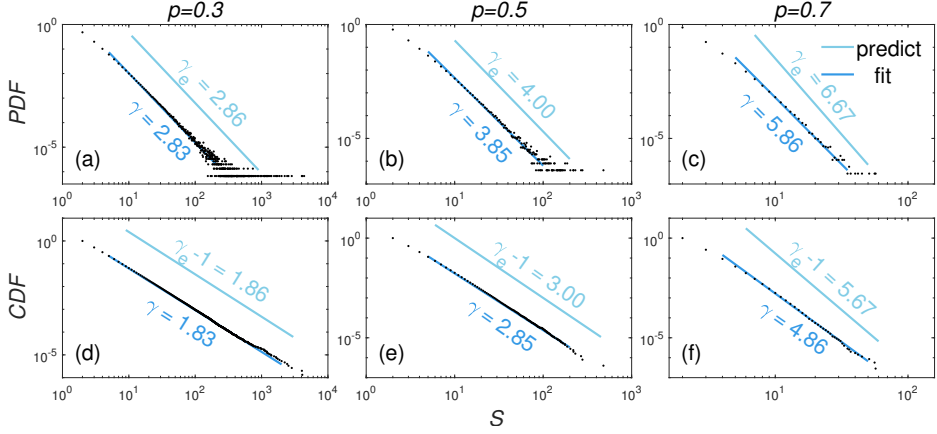}
    \caption{The size distributions of hyperedges. (a-c) are the PDF of hyperedge size distributions when $p = 0.3$, $0.5$, $0.7$, respectively. (d-f) are CDF of the same data as (a-c), respectively. The light blue line are the fitted power law by the least squares method, while the dark blue are the power law predicted by our mean-field theory given by Eq.~(\ref{gamma_e}). Parameters: $m_0=2$, $m_{new}=m_{old}=1$, and $N=5\times 10^5$.
    }
    \label{hyperedge}
\end{figure}

Fig.~\ref{powerlaw} shows that both hyperdegree and hyperedge sizes evolve to scale-invariant distributions, each showing a power law profile. Especially, for the given probability $p=0.5$, when the two evolutionary processes are equally addressed, their exponents become close to each other. These plots show that, despite its continual growth, the system organizes itself into a scale-free stationary distribution. 
In the following study, we set $N= 5\times10^5$ as the preset size and the results are averaged over 10 ensembles. Within this setup, we show the hyperdegree distribution and the size distribution of hyperedges in Fig.~\ref{hyperdegree} and Fig.~\ref{hyperedge} respectively, both with probabilities $p=0.3$, 0.5, and 0.7.

The probability density function (PDF) in Fig.~\ref{hyperdegree} (a-c) show that the network evolves into a scale-invariant distribution, where the probability that a node has $d_H$ hyperedges follows a power law, but its exponent depends on the probability $p$ instead of being constant as in the BA model. Fig.~\ref{hyperdegree} (d-f) show the corresponding cumulative distribution function (CDF) distributions, confirming the estimated power exponents in PDF profiles. As can be seen, the exponent $\gamma$ decreases (5.89, 3.86, 3.37) as the probability $p$ becomes larger, which is reasonable since the process involves more new hyperedge formation as the probability $p$ becomes larger. 

Fig.~\ref{hyperedge} shows the size distributions of hyperedges. Similarly, for the three given probabilities, the sizes of hyperedges all exhibit power law distributions, as seen in both PDF and CDF. By comparison, the exponent increases with the probability $p$. This is due to less hypergraph expansion as expected in this trend, which leads to a more homogeneous distribution.

\section{A mean-field theory}\label{theory}
To understand the evolutionary process in the DHH model, we develop a mean-field theory which adapts the idea from Ref.~\cite{Ba1999Mean-field}. We first derive the probability $P(d_H)$ and calculate the exponent for the hyperdegree distribution.
Assuming that $d_H(i)$ is continuous, any given node $i$ receives new hyperedges according to the probability $\Pi(d_H(i)) = d_H(i) / \sum_{j} d_H(j)$. 
Thus, the hyperdegree $d_H(i)$ of node $i$ approximately satisfies the following equation:
\begin{equation}
    \frac{\partial d_H(i)}{\partial t} 
    \approx p m_{old} \Pi (d_H(i))
    = p m_{old} \frac{d_H(i)}{\sum_{j} d_H(j)} .
\end{equation}

For the long-term evolution, we have $\sum_{j} d_H(j) = m_0+(m_{new}+m_{old}) p t + m_{new} (1-p) t  \approx (m_{new}+m_{old} p ) t $ and thus
\begin{equation}
    \frac{\partial d_H(i)}{\partial t} 
    \approx \frac{ m_{old} p d_H(i) }{( m_{new} + m_{old} p ) t } .
\end{equation}
Given the initial hyperdegree $d_H(i,t_i) = 1$ of the node $i$ when it joins the network, the solution of this equation is
\begin{equation}
    d_H(i,t) = (\frac{t}{t_i})^{\frac{m_{old} p} { m_{new} + m_{old}p} }.
    \label{ts_dH}
\end{equation}
With this relation between hyperdegree and joining time, the probability that a node with a hyperdegree being smaller than $d_H$ is 
\begin{align}
    P(d_H(i,t)<d_H) 
    &= P(t_i > \frac{t}{d_H^{(m_{new}+m_{old} p )/m_{old} p } })
    \notag  \\
    & = 1 - P(t_i \leqslant \frac{t}{d_H^ {(m_{new}+m_{old} p )/m_{old}p } }).
    \label{p(dht<dh)}
\end{align}
Since we assume that $m_{new}$ nodes are added to the network uniformly at each time step, we have  
\begin{equation}
    P(t_i) = \frac{1}{t}.
\end{equation}
By plugging it into Eq. \eqref{p(dht<dh)}, we have 
\begin{equation}
    P(d_H(t)<d_H) = 1 - \frac{t}{d_H^{(m_{new}+m_{old}p)/m_{old} p } t }.
\end{equation}

The hyperdegree distribution $P(d_H, t)$ of the network at time $t$ can eventually be written as
\begin{align}
    P(d_H, t) 
    & = \frac{\partial P(d_H(t)<d_H)}{\partial d_H} 
    \notag    \\
    & = \frac{m_{new}+m_{old}p}{m_{old}p}
         \  d_H^{-(2+\frac{m_{new}}{m_{old}p})} .
    \label{p(dh)}
\end{align}
Note that Eq. \eqref{p(dh)} does not include time explicitly, which indicates that the hyperdegree distribution is time-invariant. The hyperdegree distribution exhibits a power-law with an exponent 
\begin{equation}
    \gamma_d = 2+ \frac{m_{new}}{m_{old}p} \hspace{0.5em}, 
    \label{gamma_d}
\end{equation}
where $\gamma_d \in (2 , \infty)$. 

Next, we turn to the evolution of the hyperedge expansion. Similarly, by assuming the continuous change of the size $S_i$ of any given hyperedge $i$,  we have
\begin{equation}
    \frac{\partial S(i)}{\partial t}  
    \approx  m_{mew} (1-p) \Pi(S(i))
    = m_{mew} (1-p) \frac{S(i)} {\sum_j{S(j)}}.
    \label{dSize/dt}
\end{equation}
When $t$ is large enough, $\sum_j{S(j)} = (m_{new}+m_{old})pt + m_{new}(1-p)t = (m_{new}+m_{old}p) t $. So the solution to the equation \eqref{dSize/dt} is
\begin{equation}
    S(i,t) = (\frac{t}{t_i})^{\frac{ m_{new}(1-p) }{ m_{new}+m_{old}p}}.
    \label{ts_S}
\end{equation}
Following the same logic as for Eqs.(\ref{p(dht<dh)}-\ref{p(dh)}), the probability of $S$ is expressed as
\begin{equation}
    P(S) = \frac{m_{new}+m_{old}p}{m_{new}(1-p)} S^{-\gamma_e},
\end{equation}
with the exponent
\begin{equation}
    \gamma_e = 1+\frac{m_{new}+m_{old}p}{m_{new}(1-p)} \hspace{0.5em} ,
    \label{gamma_e}
\end{equation}
where $\gamma_e \in [2 , \infty)$.
In fact, by defining the ratio $ \lambda = m_{new}/m_{old} $, the temporal evolution in Eqs. \eqref{ts_dH} and \eqref{ts_S} can be rewritten as
\begin{equation}
    d_H(t) = (\frac{t}{t_i})^{\frac{ p }{ \lambda + p }} \hspace{0.5em} ,
    \hspace{0.5em}   S(t) = (\frac{t}{t_i})^{\frac{ \lambda (1-p) }{ \lambda + p }}.
    \label{lambda1}
\end{equation}
whereas the exponents can also be rewritten in a compact way as
\begin{equation}
    \gamma_d = 2+\frac{\lambda }{ p } \hspace{0.5em} ,
    \hspace{0.5em}   \gamma_e = 1+\frac{\lambda + p}{\lambda (1-p)}.
    \label{lambda2}
\end{equation}

Notably, both the temporal evolution and the two power exponents depend only on the ratio of $m_{new}$ to $m_{old}$ and the probability $p$, not on the specific values of $m_{new}$ or $m_{old}$. A slightly different analytical treatment based on the Poisson process method is also developed, with the same results as in Eqs.~(\ref{lambda1}) and~(\ref{lambda2}) being obtained. Details see Sec. II in SM~\cite{SM}.

\begin{figure}[htbp]
    \centering
    \includegraphics[width = 0.8\textwidth]{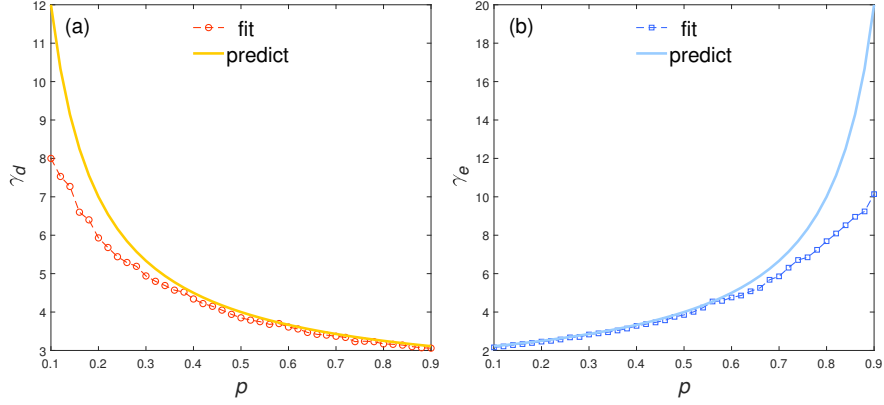}
    \caption{Comparison between the numerical estimation of power exponents and theoretical prediction for a range of probability $p$. (a) The estimated power exponent of the hyperdegree distribution (a) and the hyperedge size distribution (b). Numerical results are averaged over 20 ensembles, the solid lines are the theoretical predictions from Eq.~(\ref{lambda2}). Parameters: $m_0=2$, $m_{new}=m_{old}=1$, and $N=5\times 10^5$.
    }
    \label{comparision}
\end{figure} 

\section{Comparison}\label{comparison} 
To justify the mean-field calculation, we plot the theoretic predictions against the numerical results as in Fig.~\ref{hyperdegree} and Fig.~\ref{hyperedge} for comparison. We first see that the theoretic results are in line with the numerical results, meaning that the mean-field treatment correctly captures the dependence of the exponents on the probability $p$. However, by detailed comparison, we notice that the error becomes larger when the distribution is more skewed (with a larger exponent), as displayed in Fig.~\ref{hyperdegree}(a,d) and Fig.~\ref{hyperedge}(c,f). 

A more systematic comparison is given in Fig.~\ref{comparision}, where we plot the two estimated exponents from numerical simulations and their theoretic predictions (i.e. Eq.~(\ref{lambda2})) for a wide range of probability $p$. Fig.~\ref{comparision}(a) shows that at the higher end of $p$, the theoretical prediction is in a good agreement with the numerical estimation, but overestimates the value at the lower end. Similar observations can be made for the hyperedge size distribution in Fig.\ref{comparision} (b), but an opposite trend is displayed. The theoretic prediction is in a good agreement with a smaller $p$, while the error increases with $p$.

The opposite trend of the dependence on the probability $p$ is reasonable since the two processes of hyperedge increase and the hyperedge expansion are complementary. When $p$ is small, the preferential attachment works preferably on the hyperedge expansion. The newly joining node may thus face more choices among the existing nodes than in a pure increasing process of hyperedges as in the BA model. It turns out to dilute the ``rich-get-richer" effect regarding the hyperdegree and leads to a less heterogeneous network with a large $\gamma_d$ . 
In the opposite case when $p$ is large, the process of hyperedge increase dominates, the hyperdegrees exhibit strong heterogeneity with a smaller power exponent. As a consequence, the newly added nodes have much more hyperedges to join when they are going to contribute to the hyperedge expansion, therefore the ``rich-get-richer" effect regarding the hyperedge size is leveled down, with a less heterogeneity (i.e. a large $\gamma_e$) as expected. In fact, Eq.~(\ref{lambda2}) reveals that the two exponents are correlated with each other, i.e., $(r_e-1)(1-p)=1+1/(r_d-2)$, which vary in oppositely when $p$ is varied as we indeed observed in Fig.~\ref{comparision}.
Notice that there is an overestimation for those cases with a large exponent, one potential reason for that could be due to the fact that for those cases with less heterogeneity, there is a lack of data due to the less frequent occurrence, which may cause an enhanced error. 
  
\begin{figure}[htbp]
    \centering
    \includegraphics[width = 0.8\textwidth]{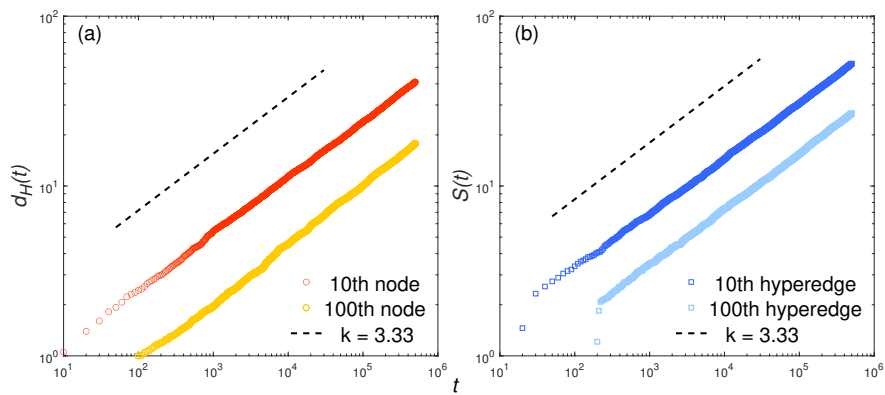}
    \caption{Typical growth of hyperdegree and hyperedge size in our DHH model with $ p=0.5 $. 
    (a) Time evolution of the hyperdegree for two nodes added to system (the 10th and 100th nodes). 
    (b) Time evolution of the hyperedge size for two hyperedges added to system (the 10th and 100th hyperedges). 
    Simulation results are obtained by averaging 100 independent runs. Both dashed lines are with the slope given by the theoretic predictions (i.e. Eq.~(\ref{lambda1})). Parameters: $m_0=2$, $m_{new}=m_{old}=1$.
    }
    \label{ts}
\end{figure}

The validation is also conducted for the temporal evolution for both the hyperdegree $d_H(t)$ or hyperedge size $S(t)$, as shown in Fig.~\ref{ts}. For the two chosen nodes, we observe that the growth profile in both cases follows power law, and the estimated power exponents in a good agreement the theoretic prediction given in Eq.~(\ref{lambda1}). Notice that, once the node is added earlier in the growth process, its advantage over the latter one remains for the whole time evolution on average [see Fig.~\ref{ts}(a)], showing the ``first-move advantage" phenomena~\cite{Newman2009first}, which are widely observable in realistic contexts. The same is also true for the hyperedge size, as seen in Fig.~\ref{ts}(b) that early formed hyperedge are generally larger than the latter one, and the advantage remains as time goes by.

\section{Conclusions and discussions} \label{conclusion}
Motivated by the gap between the observations and the state-of-the-art modeling work, we propose a simple double heterogeneity hypergraph generative model. Our model captures the coevolution of hyperedge formation and expansion, capable of showing double heterogeneous distributions, which provides a unified perspective towards many real hypernetworks. We introduce a probability that mediates two complementary processes, which leads to power-law distributions in both hyperdegree and hyperedge size. Interestingly, the scaling behaviors of the two processes are negatively correlated with each other in an intricate manner. 
We also develop a mean-field theory to grasp the essence of the evolutionary process. It gives explicit expressions for the two power exponents which are in good agreement with the numerical results. The derived temporal evolution indicates ``rich-get-richer" effect for both hyperdegree and the hyperedge size.

As a more versatile formulation, our DHH model degenerates into the classical BA model or $u$-uniform hypergraph model for some special parameter choices.
Note that, there are a few works that are able to generate non-uniform hypergraphs, such as Ref.~\cite{guo2014b}, but there the size heterogeneity of hyperedges comes from some preset distribution, while ours are time-evolving according to the law of growth. A formally similar but different model is Ref.~\cite{Courtney2017weighted}, there they propose a model of the coevolution of the hyperedge formation and weight evolution, a rich interplay between the topology and the weights is discussed.

Given the flexibility and expressive power of the hypergraph conception with regard to the classic network formulation, 
we hope this simple model is to shed more light on the understanding towards the evolution of realistic networks in a more unified way.  
 In a practical sense, we hope that our model may also provide a useful platform allowing the study of various dynamics involving high-order interactions thereon, ranging from synchronization~\cite{Carletti2023global}, game evolution~\cite{unai2021} to epidemic spreading~\cite{Iacopini2019}, among other dynamical processes~\cite{Majhi2022dynamics,Boccaletti2023structure}.

\section*{Acknowledgment}
We are supported by the National Natural Science Foundation of China under Grants Nos. 12075144 and 12165014.


\bibliographystyle{unsrt}
\bibliography{ref}
%








\end{document}